\documentclass[a4paper]{jpconf}
\usepackage[utf8]{inputenc}
\usepackage{graphicx}
\usepackage{amsmath,amssymb}
\usepackage{amsfonts}
\usepackage{physics}
\usepackage{xspace}
\usepackage{bm}
\usepackage{mathrsfs}
\usepackage{nicefrac}
\usepackage[dvipsnames]{xcolor}
\usepackage[unicode]{hyperref}
\hypersetup{colorlinks=true, citecolor=MidnightBlue, linkcolor=Blue, urlcolor=CornflowerBlue, linktocpage=true}
\usepackage[all]{hypcap}
\usepackage{multirow}
\usepackage[format=plain,justification=RaggedRight,labelsep=endash,font=small,labelfont=sc]{caption}
\usepackage[export]{adjustbox}

\definecolor {darkgreen}{rgb}{0.2,0.7,0.2}
\newcommand{\pdagger}{{\phantom{\dagger}}}
 
\begin{document}
\title{Quantum fluctuations of baryon number density}

\author{Rajeev Singh}

\address{Institute  of  Nuclear  Physics  Polish  Academy  of  Sciences,  PL-31-342  Krak\'ow,  Poland}

\ead{rajeev.singh@ifj.edu.pl}

\begin{abstract}
Quantum fluctuation expression of the baryon number for a subsystem consisting of hot relativistic spin-$\frac{1}{2}$ particles are derived. These fluctuations seems to diverge in the limit where system size goes to zero. For a broad range of thermodynamic parameters numerical solutions are obtained which might be helpful to interpret the heavy-ion experimental data.
\end{abstract}
\section{Introduction}
For many-body systems, fluctuations seems to have significant role as they provide important information about phase transition, dissipative phenomena, and clustering~\cite{Smoluchowski,PhysRevLett.85.2076,Huang:1987asp,Kubo1,Lifshitz:1963ps,PhysRevLett.49.1110}. In our recent papers~\cite{Das:2020ddr,Das:2021aar}, we studied the quantum effects and fluctuations of energy density in a hot gas of bosons and fermions. We find the increase in fluctuations for small systems. An interesting fact~\cite{Das:2020ddr,Das:2021aar} of our recent results is that even though fluctuations seem to diverge for the system going to zero size, still the fluctuations agrees with the thermodynamic limit when we sufficiently increase the system's size.

This article is based on the work~\cite{Das:2021rck}, where we studied the baryon number density fluctuations for a subsystem containing spin-$\frac{1}{2}$ particles. Current results might be important to study the QCD phase diagram and search for critical point and phenomena~\cite{Berges:1998rc,Halasz:1998qr,Stephanov:1998dy,Stephanov:1999zu,Hatta:2002sj,Son:2004iv,Stephanov:2008qz,Berdnikov:1999ph,CaronHuot:2011dr,Kitazawa:2013bta}.
Following our previous works~\cite{Das:2020ddr,Das:2021aar,Das:2021rck}, here we assume the fluctuation of the baryon number density in a small system $S_a$ of the larger thermodynamic system $S_V$ (with volume $V$), which is represented by the Grand Canonical Ensemble (GCE) and characterized by the parameters $T$ (Temperature or $\beta$ ($T^{-1}$)) and $\mu$ (Baryon Chemical Potential).

We derive the expression for the quantum fluctuations of the baryon number density operator for a hot and relativistic gas of fermions and apply this result to situations to get interesting physical insights about relativistic heavy-ion collisions. Throughout the article we assume the metric tensor with the signature $(+1,-1,-1,-1)$.  Three-vectors are denoted in bold font and a dot is used to represent the scalar product of both four and three-vectors, i.e., $a^{\mu}b_{\mu}=a\cdot b= a^0b^0-\boldsymbol{a}\cdot\boldsymbol{b}$.
\section{Basic definitions}
\label{sec:basic}
Spin-$\frac{1}{2}$ particles are represented by Dirac field operator which is given as~\cite{Tinti:2020gyh}
\begin{align}
\psi(t,\boldsymbol{x})=&\sum_r\int\frac{dK}{\sqrt{2\omega_{\boldsymbol{ k}}}}\Big(U_r^{\pdagger}(\boldsymbol{k})a_r^{\pdagger}(\boldsymbol{k})e^{-i k \cdot x}+V_r^{\pdagger}(\boldsymbol{k})b_r^{\dagger}(\boldsymbol{k})e^{i k \cdot x} \Big),
\label{equ1ver1}
\end{align}
where $dK \equiv {d^3k}/{(2\pi)^3}$, while $a_r^{\pdagger}(\boldsymbol{k})$ and $b_r^{\dagger}(\boldsymbol{k})$ represent the annihilation and creation operators for particles and antiparticles, respectively, whereas $r$ denotes the polarization degree of freedom.
Operators $a_r^{\pdagger}(\boldsymbol{k})$ and $b_r^{\dagger}(\boldsymbol{k})$ satisfy the following canonical anticommutation rules, $\{a_r^{\pdagger}(\boldsymbol{k}),a_s^{\dagger}(\boldsymbol{k}^{\prime})\} =(2\pi)^3\delta_{rs} \delta^{(3)}(\boldsymbol{k}-\boldsymbol{k}^{\prime})$ and $ \{b_r^{\pdagger}(\boldsymbol{k}),b_s^{\dagger}(\boldsymbol{k}^{\prime})\} =(2\pi)^3\delta_{rs}\delta^{(3)}(\boldsymbol{k}-\boldsymbol{k}^{\prime})$. $U_r^{\pdagger}(\boldsymbol{k})$ and $V_r^{\pdagger}(\boldsymbol{k})$ are called Dirac spinors which have normalization ${\bar U}_r^{\pdagger}(\boldsymbol{k}) U_s^{\pdagger}(\boldsymbol{k}) = 2 m \delta_{rs}$ and ${\bar V}_r^{\pdagger}(\boldsymbol{k}) V_s^{\pdagger}(\boldsymbol{k}) = -2 m \delta_{rs}$, and $\omega_{\boldsymbol{k}}=\sqrt{\boldsymbol{k}^2+m^2}$ is the particle's energy.

To calculate thermal averaging of quantum operators we need to evaluate the thermal expectation values of the products of two and four creation and/or annihilation operators (for both particles and antiparticles)~\cite{CohenTannoudji:422962,Itzykson:1980rh,Evans:1996bha}
\begin{align}
\langle a_r^{\dagger}({\boldsymbol{k}})a_s^{\pdagger}({\boldsymbol{k}}^{\prime})\rangle &=(2\pi)^3\delta_{rs}\delta^{(3)}({\boldsymbol{k}}-{\boldsymbol{k}}^{\prime})f(\omega_{\boldsymbol{k}}),\label{equ2ver1}\\
\langle a^{\dagger}_r(\boldsymbol{k})a^{\dagger}_s(\boldsymbol{k}^{\prime})a_{r^{\prime}}^{\pdagger}(\boldsymbol{p})a_{s^{\prime}}^{\pdagger}(\boldsymbol{p}^{\prime})\rangle &=(2\pi)^6 \Big(\delta_{rs^{\prime}}\delta_{r^{\prime}s}\delta^{(3)}(\boldsymbol{k}-\boldsymbol{p}^{\prime})~\delta^{(3)}(\boldsymbol{k}^{\prime}-\boldsymbol{p})\nonumber\\
&-\delta_{rr^{\prime}}\delta_{ss^{\prime}}\delta^{(3)}({\boldsymbol{k}}-\boldsymbol{p})~\delta^{(3)}({\boldsymbol{k}}^{\prime}-\boldsymbol{p}^{\prime})\Big)f(\omega_{{\boldsymbol{k}}})f(\omega_{{\boldsymbol{k}}^{\prime}}).\label{equ3ver1}
\end{align}
where $f(\omega_{{\boldsymbol{k}}})=1/(\exp(\beta(\omega_{\boldsymbol{k}}-\mu))+1)$ is the distribution function for particles and similarly, $\bar{f}(\omega_{{\boldsymbol{k}}})=1/(\exp(\beta(\omega_{\boldsymbol{k}}+\mu))+1)$ for antiparticles.

We define baryon number density~\cite{Chen:2018cts} operator $\hat{J}^{0}_a$ associated with the conserved baryon current in a subsystem $S_a$ as
\begin{align}
\hat{J}^{0}_a = \frac{1}{(a\sqrt{\pi})^3}\int d^3\boldsymbol{x}~\hat{J}^{0}(x)~\exp\left(-\frac{{\boldsymbol{x}}^2}{a^2}\right),
\label{equ4ver1}
\end{align}
where $\hat{J}^{0} = \psi^{\dagger}\psi$, and we use a smooth Gaussian profile with a length scale $a$ placed at the origin of the coordinate system in order to avoid sharp-boundary effects.
Following expression for variance is used to calculate the fluctuation for baryon number of the subsystem $S_a$
\begin{equation}
 \sigma^2(a,m,T,\mu) = \langle :\hat{J}^{0}_a: :\hat{J}^{0}_a: \rangle - \langle :\hat{J}^{0}_a :\rangle^2\, 
 \label{equ5ver1}
\end{equation}
and the normalized standard deviation is expressed as 
\begin{equation} 
\sigma_n(a,m,T,\mu)= \frac{(\langle:\hat{J}^{0}_a::\hat{J}^{0}_a:\rangle- \langle :\hat{J}^{0}_a :\rangle^2)^{1/2}}{\langle :\hat{J}^{0}_a :\rangle}.
\label{equ6ver1}
\end{equation}
where $\langle :\hat{J}^{0}_a :\rangle$ is the thermal expectation value of the normal ordered operator $:\hat{J}^{0}_a:$, we use procedure of normal ordering in order to remove an infinite vacuum part which usually comes from zero-point energy contributions. 
\section{Quantum fluctuation expression}
\label{sec:fluctuation}
Using Equation~\eqref{equ2ver1}, the thermal expectation value of $:\hat{J}_a^0:$ is given as the following which agrees with the kinetic theory definition
\begin{align}
    \langle :\hat{J}_a^0: \rangle = 2\int dK \Big[f(\omega_{\boldsymbol{k}})-\bar{f}(\omega_{\boldsymbol{k}})\Big]. 
    \label{equ8ver1}
\end{align}
here, factor 2 accounts for degeneracy in spin.

Again using Equation~\eqref{equ3ver1} we obtain the variance to determine the fluctuation for the baryon number in the subsystem $S_a$
\begin{align}
    \sigma^2(a,m,T,\mu)&=\int \frac{dK}{\omega_{\boldsymbol{k}}}\frac{dK^\prime}{\omega_{\boldsymbol{k}^{\prime}}}
    (\omega_{\boldsymbol{k}}\omega_{\boldsymbol{k}^{\prime}}+\boldsymbol{k}\cdot\boldsymbol{k}^{\prime}+m^2) e^{-\frac{a^2}{2}(\boldsymbol{k}-\boldsymbol{k}^{\prime})^2}\times\nonumber\\
    &~~~~~~~~~~\left[f(\omega_{\boldsymbol{k}})\left(1-f(\omega_{\boldsymbol{k}^{\prime}})\right)+\bar{f}(\omega_{\boldsymbol{k}})(1-\bar{f}(\omega_{\boldsymbol{k}^{\prime}}))\right]\nonumber\\
    & -\int \frac{dK}{\omega_{\boldsymbol{k}}}\frac{dK^\prime}{\omega_{\boldsymbol{k}^{\prime}}}
    (\omega_{\boldsymbol{k}}\omega_{\boldsymbol{k}^{\prime}}+\boldsymbol{k}\cdot\boldsymbol{k}^{\prime}-m^2) e^{-\frac{a^2}{2}(\boldsymbol{k}+\boldsymbol{k}^{\prime})^2}\times\nonumber\\
    &~~~~~~~~~~\left[f(\omega_{\boldsymbol{k}})(1-\bar{f}(\omega_{\boldsymbol{k}^{\prime}}))+\bar{f}(\omega_{\boldsymbol{k}})(1-f(\omega_{\boldsymbol{k}^{\prime}}))\right].
    \label{equ9ver1}
\end{align}
In Equation~\eqref{equ9ver1}, which by the way is our main result, we discard the divergent temperature and baryon chemical potential independent vacuum term~\cite{Das:2020ddr,Das:2021aar,Das:2021rck}. Since the spin and particle-antiparticle degrees of freedom are already included in above expression, therefore to take into account the degeneracy factors $(g)$ due to other internal degrees of freedom, we need to do the following substitutions: $\langle :\hat{J}_a^0: \rangle\rightarrow g \langle :\hat{J}_a^0: \rangle$ and $\sigma^2\rightarrow g \sigma^2$~\cite{Das:2020ddr,Das:2021aar,Das:2021rck}.  
\begin{figure}[t]
	\includegraphics[scale=0.44]{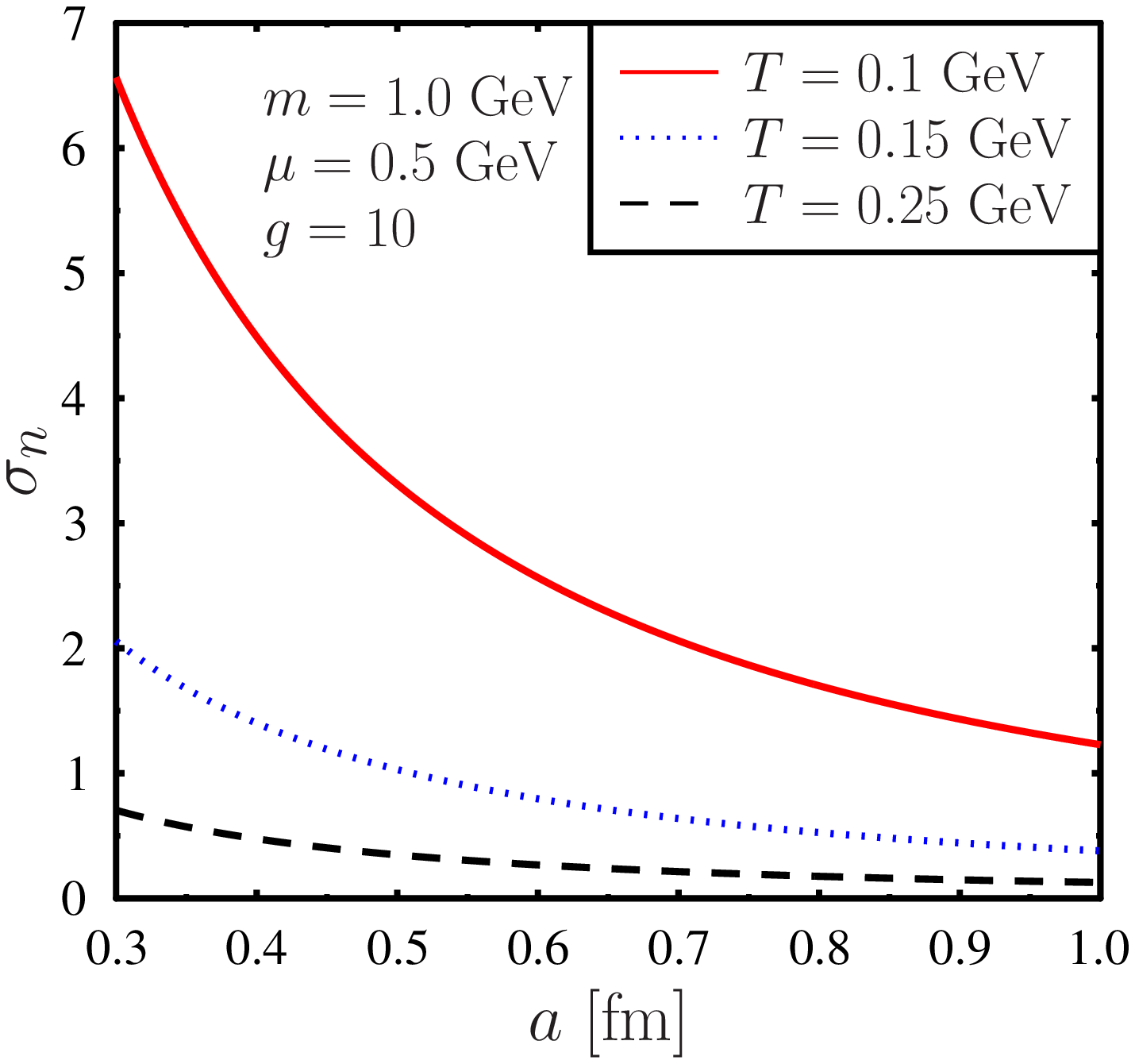}
	\includegraphics[scale=0.44]{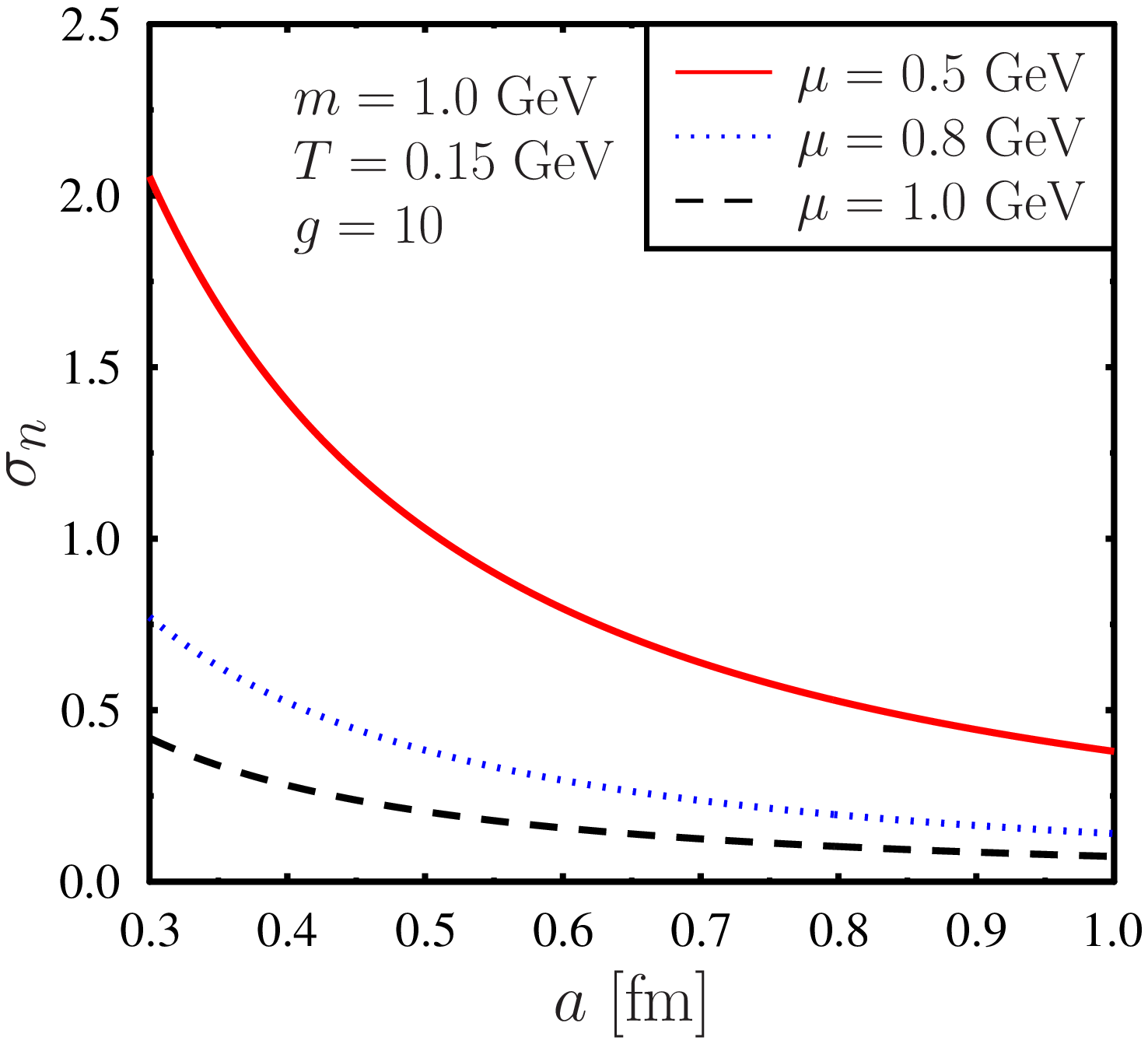}
	\includegraphics[scale=0.44]{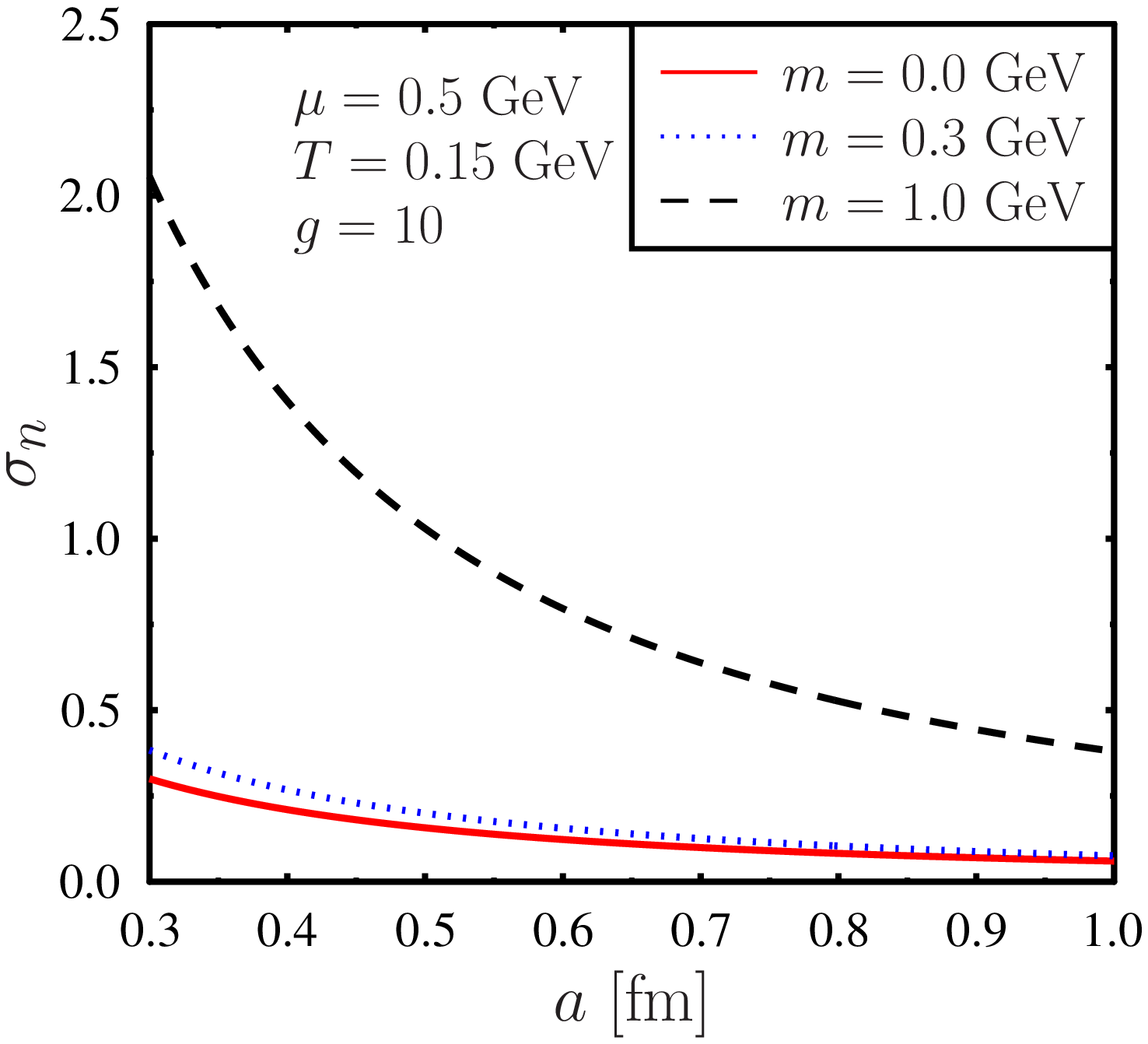}
	\caption{Variation of normalized fluctuation $\sigma_n$ in the subsystem $S_a$ with the scale $a$ (Top left) for different values of the temperature $T$ with fixed particle mass $m$ and baryon chemical potential $\mu$.
	(Top right) for different values of the baryon chemical potential $\mu$ with fixed particle mass $m$ and temperature $T$.
	(Bottom) for different values of the particle mass with fixed temperature $T$ and baryon chemical potential $\mu$.}
	\label{fig:1}
\end{figure}
\section{Thermodynamic limit}
\label{sec:thermo}
Since $S_a$ is a subsystem of the the larger thermodynamic system $S_V$, the thermodynamic limit can be obtained by by considering the $a\rightarrow \infty$ limit, in this limit, our quantum fluctuation reduces to the formula similar to the classical statistical fluctuation. Hence to get the thermodynamic limit we need to study the susceptibilities which describe the fluctuations in the baryon number obtained for thermo-chemical equilibrium~\cite{Nahrgang:2014fza}
\begin{align}
    \chi_l^{(B)}=\frac{\partial^l(P/T^4)}{\partial(\mu/T)^l}\bigg\vert_T.
    \label{equ10ver1}
\end{align}
where $P$ represents the thermodynamic pressure. Susceptibilities can also be represented by the cumulants of the baryon distribution with $n_B\equiv N_B/V$ being the net baryon number density, as,
\begin{align}
    & \chi_1^{(B)}=\frac{1}{VT^3}\langle N_B\rangle=\frac{1}{T^3}\frac{\langle N_B\rangle}{V}=\frac{n_B}{T^3},\nonumber\\
    & \chi_2^{(B)}=\frac{1}{VT^3}\langle (\triangle N_B)^2\rangle=\frac{1}{VT^3}\langle (N_B-\langle N_B\rangle)^2 \rangle=\frac{V}{T^3} \langle (n_B-\langle n_B\rangle)^2 \rangle.
    \label{equ11ver1}
\end{align}
%
%
The second order susceptibility $\chi_2^{(B)}$ can again be evaluated by taking the derivative of the thermodynamic pressure using Eq.~\eqref{equ10ver1}, where at finite temperature and baryon chemical potential, the thermodynamic pressure is given as~\cite{Nahrgang:2014fza}
\begin{align}
    \frac{P}{T^4} = & \frac{2g}{T^3}\int dK\Bigg[ \ln\bigg(1+\exp\left(-\frac{\omega_{\boldsymbol{k}}-\mu}{T}\right)\bigg) +\ln\bigg(1+\exp\left(-\frac{\omega_{\boldsymbol{k}}+\mu}{T}\right)\bigg)\Bigg]. 
    \label{equ13ver1}
\end{align}
By making use of Eqs.~\eqref{equ10ver1} and \eqref{equ13ver1} we can easily come to the following expressions 
\begin{align}
  \chi_1^{(B)} &=  \frac{\partial(P/T^4)}{\partial(\mu/T)}\bigg\vert_T = \frac{2g}{T^3}\int dK \Big[f(\omega_{\boldsymbol{k}})-\bar{f}(\omega_{\boldsymbol{k}})\Big] = \frac{n_B}{T^3}
  \label{equ14ver1}
\end{align}
and,
\begin{align}
    \chi_2^{(B)} &=  \frac{\partial^2(P/T^4)}{\partial(\mu/T)^2}\bigg\vert_T = \frac{2g}{T^3} \int dK \Big[f(\omega_{\boldsymbol{k}})(1-f(\omega_{\boldsymbol{k}}))+\bar{f}(\omega_{\boldsymbol{k}})(1-\bar{f}(\omega_{\boldsymbol{k}}))\Big].
    \label{equ15ver1}
\end{align}
Hence, we arrive at
\begin{align}
    T^3\chi_2^{(B)}&=V \langle (n_B-\langle n_B\rangle)^2 \rangle = 2g\int dK \Big[f(\omega_{\boldsymbol{k}})(1-f(\omega_{\boldsymbol{k}}))+\bar{f}(\omega_{\boldsymbol{k}})(1-\bar{f}(\omega_{\boldsymbol{k}}))\Big].
    \label{equ16ver1}
\end{align}
For the case of large volume limit, Eq.~\eqref{equ9ver1} should reproduce to Eq.~\eqref{equ16ver1}, which can be seen by using the following Gaussian representation of the 3D Dirac delta function
\begin{align}
    \delta^{(3)}({\boldsymbol{k}}-{\boldsymbol{k}^\prime})=\lim_{a \to\infty} \frac{a^3}{(2\pi)^{3/2}}e^{-\frac{a^2}{2}({\boldsymbol{k}}-{\boldsymbol{k}^\prime})^2}.
    \label{equ17ver1}
\end{align}
which leads to
\begin{align}
     \lim_{a\to\infty} a^3(2\pi)^{3/2}\bigg[\langle :\hat{J}_a^0: :\hat{J}_a^0:\rangle-\langle :\hat{J}_a^0: \rangle^2\bigg] = 2\int dK \Big[f(\omega_{\boldsymbol{k}})(1-f(\omega_{\boldsymbol{k}}))+\bar{f}(\omega_{\boldsymbol{k}})(1-\bar{f}(\omega_{\boldsymbol{k}}))\Big],
    \label{equ18ver1}
\end{align}
Thus,
\begin{align}
     \lim_{a\to\infty} V_a \bigg[\langle :\hat{J}_a^0: :\hat{J}_a^0:\rangle-\langle :\hat{J}_a^0: \rangle^2\bigg]=T^3\chi_2^{(B)},
    \label{equ19ver1}
\end{align}
where $V_a = a^3 (2\pi)^{3/2}$ can be called the volume of the ``Gaussian'' subsystem $S_a$.

\begin{figure}[t]
	\includegraphics[scale=0.44]{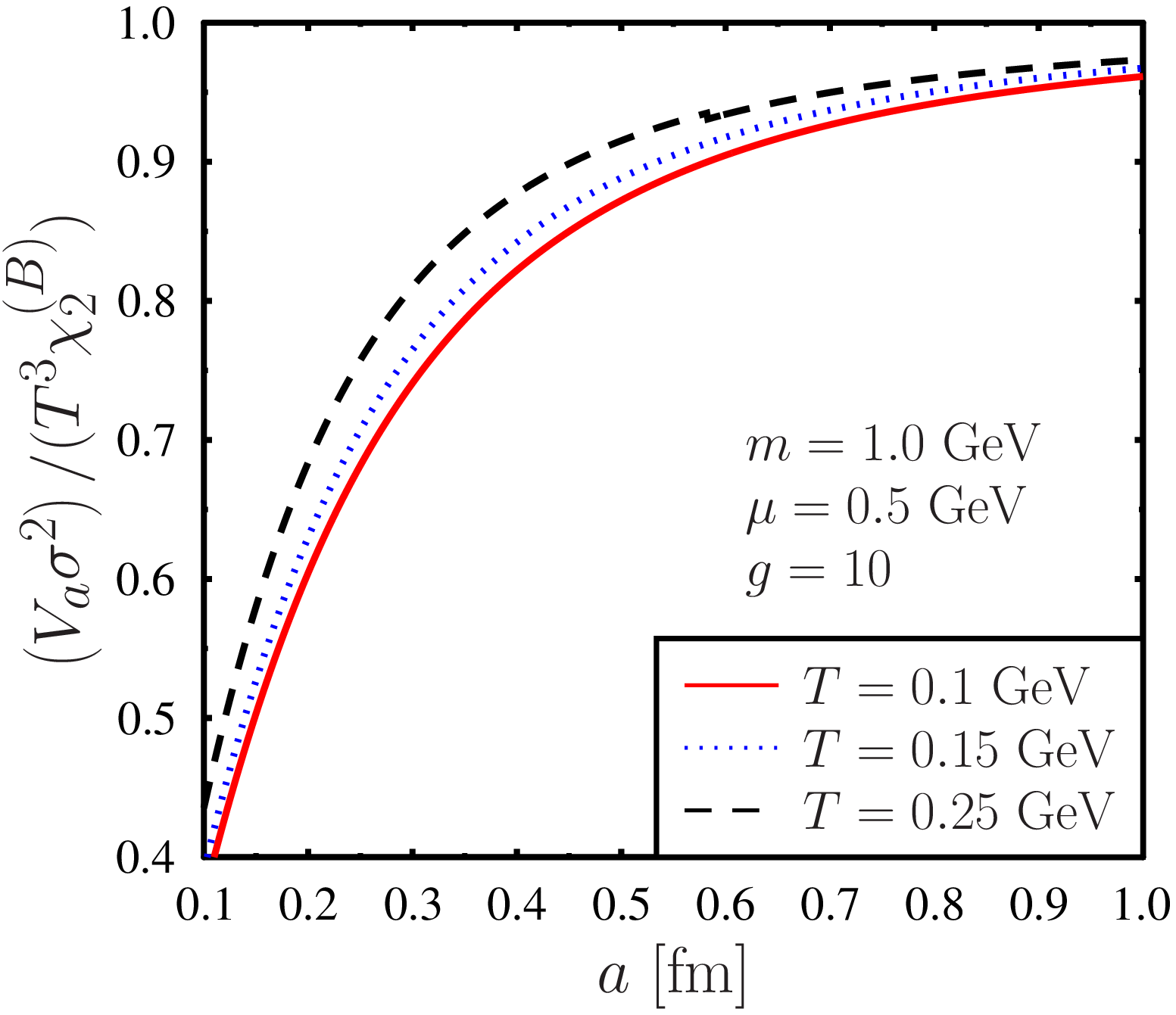}
	\includegraphics[scale=0.44]{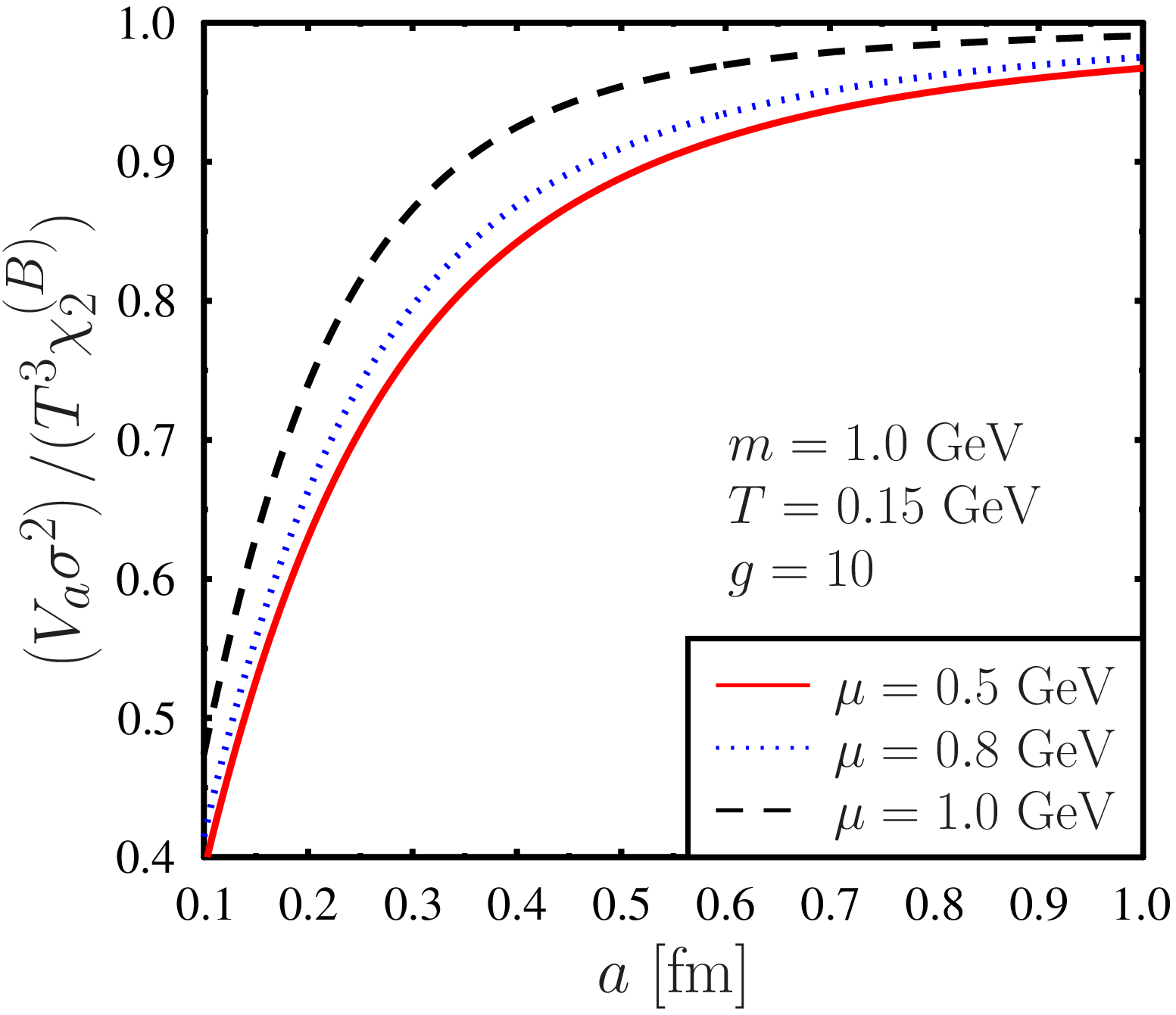}
	\includegraphics[scale=0.44]{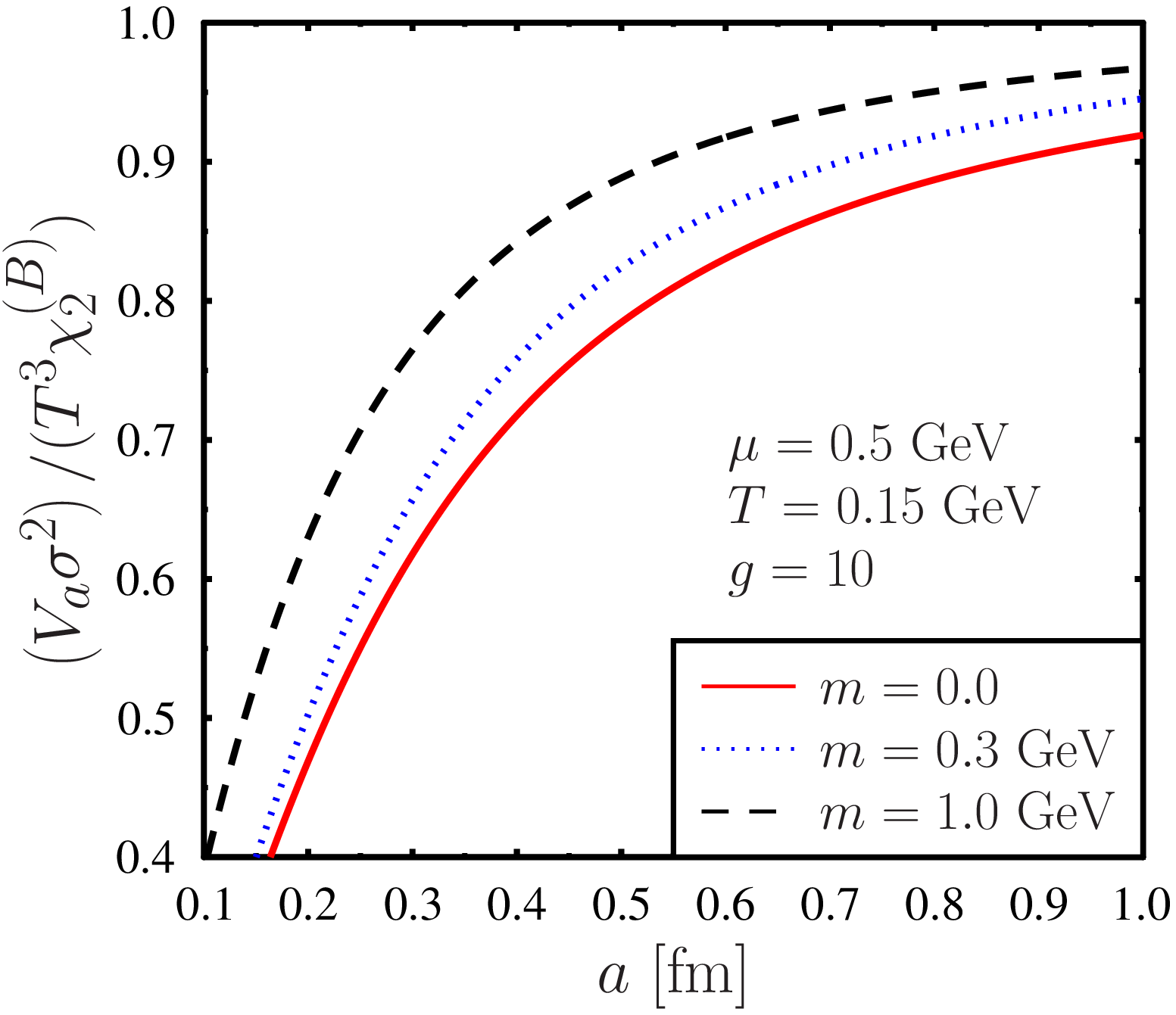}
	\caption{Variation of normalized fluctuation $V_a\sigma^2/(T^3\chi_2^{(B)})$ (Top left) for different values of temperature ($T$) with fixed baryon chemical potential ($\mu$) and particle mass ($m$).
	(Top right) for different values of baryon chemical potential ($\mu$) with fixed temperature ($T$) and particle mass ($m$).
	(Bottom) for different values of particle mass ($m$) with fixed temperature ($T$) and baryon chemical potential ($\mu$).}
	\label{fig:4}
\end{figure}
\section{Numerical results}
\label{sec:numer}
Our main finding is given by Eq.~\eqref{equ9ver1} which describe fluctuations of the baryon number density. Using simple numerical integration and for a broad range of thermodynamic parameters ($0.1$ GeV $\leq T \leq 0.25$ GeV, $0.5$ GeV $\leq \mu \leq 1.0$ GeV and $0.0$ $\leq m \leq 1.0$ GeV), we can get results for any subsystem size $a$. Figs.~\eqref{fig:1} show the variation of the normalized fluctuation $\sigma_n$ with respect to subsystem size $a$ on the $x$-axis for different values of temperature, baryon chemical potential and particle mass, respectively, with the internal degeneracy factor to be $g= 10$.

From Figs.~\eqref{fig:1}, we saw that the normalized fluctuation $\sigma_n$ decreases with the increase in the subsystem size $S_a$, but for small system size, the fluctuation increases drastically, which is due to the quantum mechanical behavior. These plots indicate that $\sigma_n$ decreases with the increase in temperature and baryon chemical potential, and increases with the increase in the particle mass.

$V_a\sigma^2/(T^3\chi_2^{(B)})$ should approach one in the $a\rightarrow \infty$ limit (i.e., thermodynamic limit) as can be easily seen from Eq.~\eqref{equ19ver1}, which is demonstrated by Figs.~\eqref{fig:4}, again plotted for broad range of values of temperature, baryon chemical potential and mass of the particle, respectively.
\section{Conclusion}
\label{sec:conc}
In this paper, we have presented the quantum fluctuations in baryon number in subsystems of a hot and dense relativistic gas of spin-$\frac{1}{2}$ particles. Even though our results seem to diverge for the system going to zero size, they still agree with the results known from statistical physics for the system going to sufficiently large size $a$. The numerical findings obtained here might be useful to interpret the heavy-ion experimental data differently and give new information. 
\medskip
\section*{Acknowledgments}
I am very thankful to A. Das, W. Florkowski, and R. Ryblewski for their fruitful collaboration. This research was supported in part by the Polish National Science Centre Grants No. 2016/23/B/ST2/00717 and No. 2018/30/E/ST2/00432, and IFJ PAN.
\section*{References}
\bibliographystyle{iopart-num}
\bibliography{fluctuationRef.bib}{}
\end{document}